# Phase diagram and superconductivity of polonium hydrides under high pressure

Yunxian Liu, Defang Duan, Fubo Tian, Da Li, Xiaojing Sha, Zhonglong Zhao, Huadi Zhang, Gang Wu, Hongyu Yu, Bingbing Liu, & Tian Cui

*State Key Laboratory of Superhard Materials, College of Physics, Jilin University, Changchun 130012, People's Republic of China*

*Corresponding Author: E-mail: cuitian@jlu.edu.cn

## Abstract

High pressure structures, phase diagram and superconductivity of polonium hydrides have been systematically investigated through the first-principles calculations based on the density functional theory. With the increasing pressure, several stoichiometries (PoH, $PoH_2$, $PoH_4$ and $PoH_6$) are predicted to stabilize in the excess hydrogen environment. All of the reported hydrides, exception of PoH, exhibit intriguing structural character with the appearing $H_2$ units. Moreover, the electronic band structure and projected density of states (PDOS) demonstrate that these energetically stable phases are metallic. The application of the Allen-Dynes modified McMillan equation with the calculated electron-phonon coupling parameter reveals that $PoH_4$ is a superconductor with a critical temperature $T_c$ of 41.2-47.2 K at 300 GPa.

## Introduction

Over the past ten years, hydrogen-rich compounds have been the focus of research for their potential to become extraordinary high-temperature superconductor at lower pressures accessible to current experiments. The elements of group VIA, which include oxygen, sulfur, selenium, tellurium and polonium elements, present many interesting properties under pressure, such as the metallic phases of $O_2$, S, Se and Te are found to be superconducting in experiment, with a transition temperature $T_c$ of 0.6, 10, 5 and 4.5 K,[1-5] respectively. Recently, group VIA hydrides are expected to metallize at lower pressures with potential superconductivity. Remarkably, the high-pressure ordered crystal structures of $(H_2S)_2H_2$ ($H_3S$ with a H:S stoichiometric ratio of 3:1) were first predicted by our group using the *ab initio* calculation method.[6] And the cubic phase (space group *Im-3m*) becomes a high-temperature superconductor with outstanding high $T_c$ of 191~204 K at 200 GPa. Subsequently, the superconductivity in $H_2S$ sample with a transition temperature ($T_c$) of 190 K above 150 GPa was observed by Eremets et al in experiment.[7] Later, theoretical studies by our group have reported that $H_2S$ really decomposes into S and $H_3S$ above 50 GPa, and $H_3S$ is stable at least up to 300 GPa, which proved that high $T_c$ in $H_2S$ sample observed experimentally comes from $H_3S$.[8] Other theoretical studies also authenticated our results using different structural search techniques.[9-10] Recently, selenium (Se) hydrides and tellurium (Te) hydrides were theoretically predicted to exhibit $T_c$ in the range of 40-131 K and 46-104 K at megabar pressures, respectively.[11-13] These dense hydrides has reignited great interest in chalcogen hydrogen-rich compounds.

Polonium (Po) belongs to the chalcogenide family of the periodic table, which is the only metal element in the group VIA. In addition, Po is one of the strangest elements of the periodic table, which possesses the simple cubic (sc) structure at ambient conditions, while most elements are tendency to form complex crystal structures (fcc, bcc, hcp, etc). Based on the intriguing nature of Po, H-rich polonium hydrides might exhibit different properties over other chalcogenide hydrides. It is well established that the application of high pressure can influence the physical and chemical properties of matter, and lead to the formation of unprecedented phases or complexes that have novel characters. Such as, the superconductivity of $SiH_4$ and PtH was confirmed under pressure experimentally.[14-15] Other new unusual stoichiometries was predicted in theory (e.g. $LiH_n$, $NaH_n$, $KH_n$, $RbH_n$, $CsH_n$ and so on).[16-20] Therefore, the investigation of polonium hydrides under high pressure might provide us with interesting information.

In this article, to explore the ground-state structures of Po-H system over a range of pressures, we use the recently fast-developed evolutionary algorithm Universal Structure Predictor: Evolutionary Xtallography (USPEX)[21-23] Moreover, we investigate in detail the optimum static structures as a function of pressure, their dynamical stability, the corresponding electronic band structures and superconductivity of Po-H. Our results show that exception of PoH, other stable stoichiometry phase appear $H_2$ units. Moreover, all the stable polonium hydrides are found metallic, and the superconducting critical temperature $T_c$ values of PoH, $PoH_4$, and $PoH_6$ are 0.14-0.65 K, 41.1-47.2 K, and 2.65-4.68K at 300, 200 and 200 GPa, respectively.

## Computational Methods

The searches for stable high pressure structures of the $PoH_n$ (n = 1-6) system were performed through the evolutionary algorithm, implemented in the USPEX code.[21-23] Structure predictions are implemented at 50, 100, 200, and 300 GPa with 1-4 formula units (f.u.). Each structure was fully relaxed to an energy minimum using density functional theory with the Perdew–Burke–Ernzerhof (PBE) form of the generalized gradient approximation (GGA)[24] implemented in the Vienna *ab initio* simulation package VASP code.[25] We optimized predicted stable structures at higher accurate level. A plane-wave basis set cutoff of 800 eV and a Brillouin zone sampling grid of spacing $2\pi \times 0.03$ Å$^{-1}$ were chose, which ensures total energy convergence better than 1 meV/atom. The all-electron projector augmented wave method (PAW)[26] is adopted with the cut-off radius of 0.8 a.u. and 2.1 a.u. for H ($1s^2$) and Po ($6s^2 6p^4$), respectively. In the geometrical optimization, all forces on atoms were converged to less than 0.005eV/Å. The lattice dynamics and electron-phonon coupling electron-phonon coupling (EPC) for superconducting properties of stable compounds have been computed with QUANTUM-ESPRESSO.[27] Convergence tests provide a suitable value of 80 Ry kinetic energy cutoff. The q-point mesh in the first BZ of 4×4×2 for $P6_3/mmc$ (PoH), 4×4×2 for $C2/c$ ($PoH_4$), and 3×3×3 for $C2/m$ ($PoH_6$) structures are used in the Brillouin zone.

## Results and discussion

We explore polonium hydrides with hydrogen-rich contents through the evolutionary algorithm USPEX. The calculations were performed at 50-300 GPa with considering simulation sizes ranging from one to four formula units (f.u.) for $PoH_n$ (n=1, 2, 3, 4, 5, 6). Our calculations gave convex hulls on the formation enthalpy of the Po-H system with H-rich stoichiometry from 50 to 300 GPa shown in Figure 1a. The phases on the convex hull are thermodynamically stable and may be experimentally synthesized in principle, whereas those above it are metastable. The essential information can be summarized as follows: (i) at 100 GPa, $PoH_2$ had the most negative enthalpy out of all the structures that we examined; (ii) when pressure is up to 150 GPa, the most stable phase is still $PoH_2$, with $PoH_4$ falling on the convex hull. (iii) at 200 GPa, $PoH_2$, $PoH_4$ and $PoH_6$ stoichiometries were all found to lie on the hull (vi) With increasing pressure to 250 GPa, another stable stoichiometry of PoH appears on the convex hull and it emerges as the most stable phase up to 300 GPa. (v) At 300 GPa, except for $PoH_3$ and $PoH_5$, all the stoichiometries become stabilized on the hull. Moreover, the stable pressure range for the predicted $PoH_n$ stable structures with Po and $H_2$[28-29] were given at the selected pressures. Figure 1b present the pressure-composition phase diagram of the Po-H system. It shows the stability pressure ranges of PoH, $PoH_2$, $PoH_4$ and $PoH_6$. The *Cmcm* $PoH_2$ becomes stable at 84.5 GPa, when up to 133 GPa, the *Pnma* becomes energetically favorable. For $PoH_4$ (space group $C2/c$), $PoH_6$ (space group $C2/m$) and PoH (space group $P6_3/mmc$) become stable at 137, 195.5 and 208

GPa, respectively.

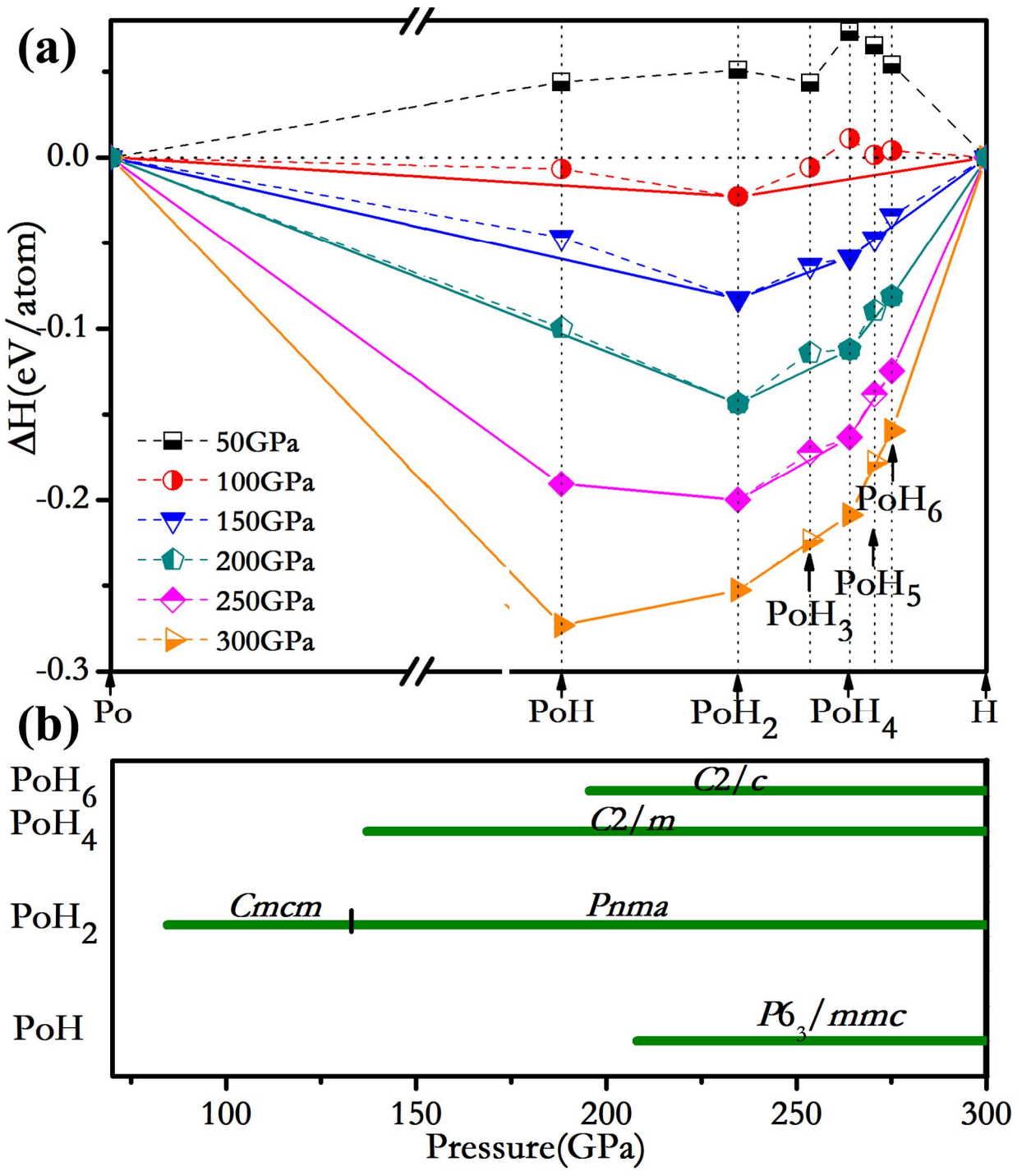

**Figure 1** (a) The enthalpies of formation ($\Delta H_f$, in eV/atom) for Po-H system with respect to Po and H at 50, 100, 150, 200, 250 and 300 GPa. (b) The phase diagram of Po-H system in the pressure range from 70 to 300 GPa.

The crystal structures motifs of stable stoichiometries are depicted in Fig. 2, which shows that all structures appear $H_2$ units exception of PoH. From our calculations, the predicted energetically favored structure of PoH is *P*6$_3$/*mmc* at 300 GPa with H atoms sitting at the 2c and 2a Wyckoff site, as depicted in Fig. 2a. The predicted crystalline phase of $PoH_2$ has two structures in the *Cmcm* and *Pnma* space group at 50 and 300 GPa, respectively. For *Cmcm* structure, Po forms three-dimensional network, which is consist with a tetrahedron and half of a octahedron (Fig. 2b). While in *Pnma* phase, Po atoms forms a distorted bcc lattice (Fig. 2c). Moreover, quasi-molecular $H_2$-unit were found with the distance of H-H 0.793 Å (50 GPa in *Cmcm* ) and 0.841 Å (200 GPa in *Pnma*), respectively. For $PoH_4$ stoichiometry (Fig. 2d), it posses a *C*2/*c* symmetry at 200 GPa, where Po atoms built up 2D layers and the $H_2$ units have a H-H distance of 0.818 Å. Turning to the richest-$H_2$ stoichiometry $PoH_6$, it has a monoclinic *C*2/*m* space group at 200 GPa with chain Po and $H_2$ units, as shown in Fig. 2e. Calculated structural parameters of the predicted stable structures for Po-H compounds at the selected pressure are displayed in Table S1.

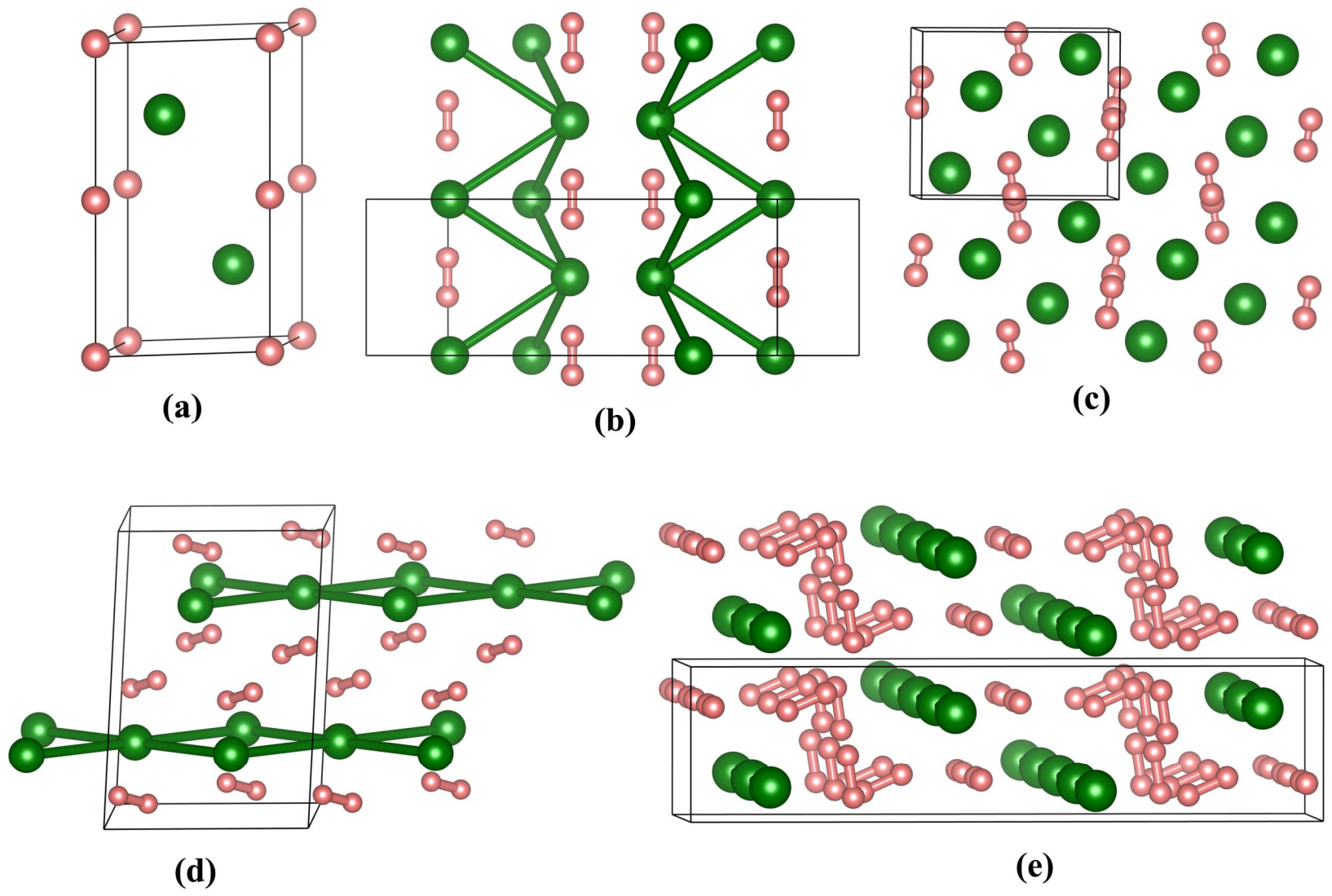

**Figure 2** The selected stable phases for Po-H system. Green atoms depict Po, while pink atoms present H, (a) PoH-$P6_3/mmc$ at 300 GPa, (b) $PoH_2$-*Cmcm* at 50 GPa, (c) $PoH_2$-*Pnma* at 300 GPa, (d) $PoH_4$-*C*2/*c* at 200 GPa and (e) $PoH_6$-*C*2/*m* at 200 GPa.

To determine the dynamical stability of the thermodynamically stable structures is necessary. Figure S1 presents the calculated phonon band dispersions and projected phonon density of states (PHDOS) for $P6_3/mmc$ PoH, *Pnma* $PoH_2$, *C*2/*c* $PoH_4$, and *C*2/*m* $PoH_6$. They are all dynamical stability in their accessible pressures by the evidence of the absence of any imaginary phonon mode in the entire Brillouin zone. Low energy phonon modes are mainly from Po atoms owing to their much high atomic mass, whereas modes with high frequency region are associated with H atoms.

Band structure and projected density of states (DOS) at different pressures for PoH, $PoH_2$, $PoH_4$ and $PoH_6$ in the $P6_3/mmc$, *Pnma*, *C*2/*c* and *C*2/*m* are calculated in order to analysis the electronic properties, as displayed in Figure 3. Clearly, the predicted structures exhibit conductor property with overlap between the conduction and the valence bands under pressure. The metallic behavior of Po-H system indicates that these phase might be superconductors and we will discuss it in the following. For the simplest stoichiometry PoH-$P6_3/mmc$ structure and $PoH_2$-*Pnma*, the Po-p state predominate the energy range from 5eV to -10 eV, and, there exist strong hybridization between Po-p and H-s states (-10 to -15 eV), as can be seen in Figure (3a, 3b). In $PoH_4$, we can see that around the Fermi level the domination are Po-s Po-p and H-s orbits (Fig. 3c). For H-rich $C2/m$ of $PoH_6$, at the Fermi level a strong DOS peak was observed in Figure 5d, which originate from Po-p and H-s electrons.

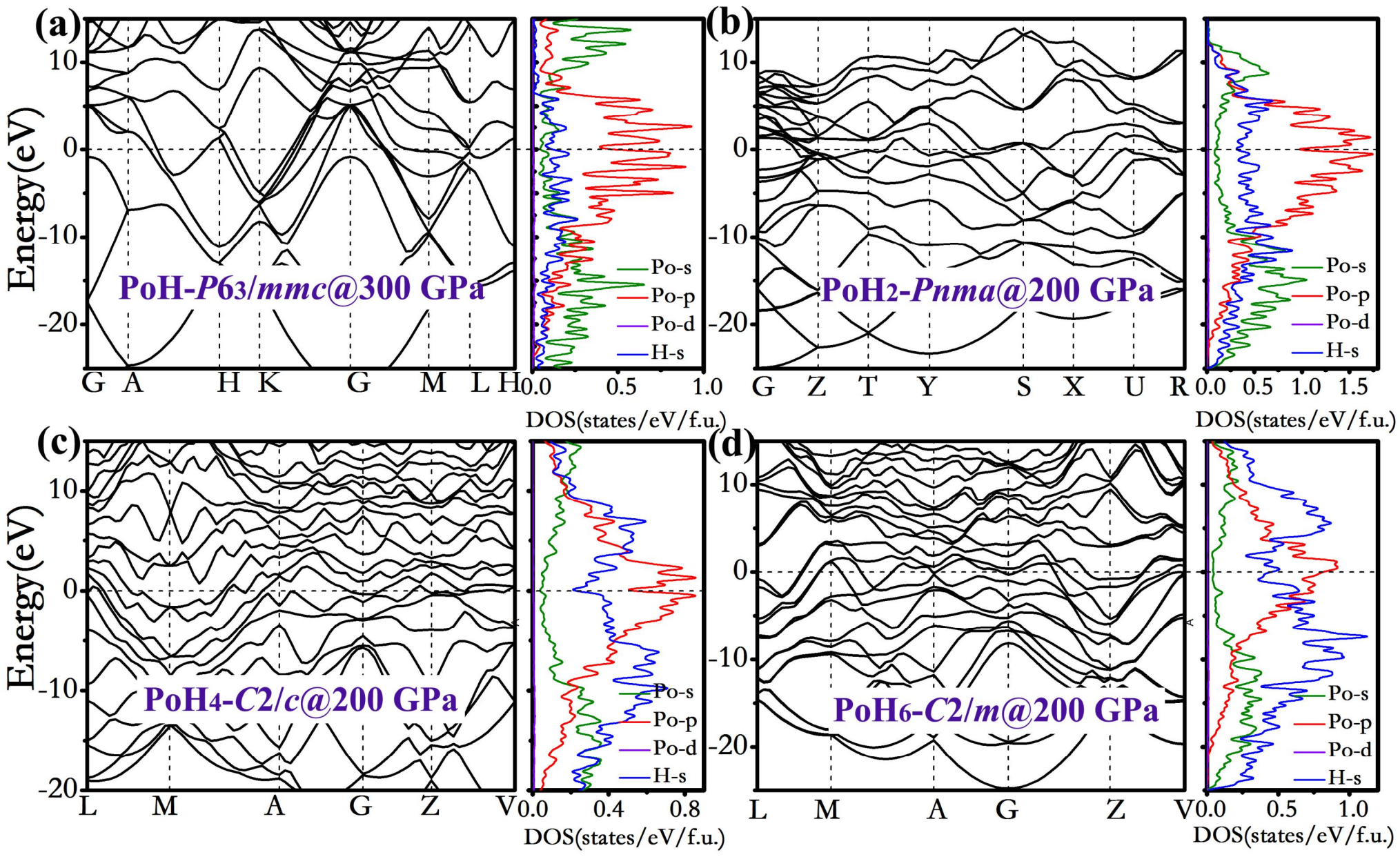

**Figure 3** Electronic band structure and partial density of states (PDOS). (a) PoH-$P6_3/mmc$ at 300 GPa, (b) $PoH_2$-*Pnma* at 200 GPa, (c) $PoH_4$-$C2/c$ at 200 GPa and (d) $PoH_6$-$C2/m$ at 200 GPa.

We investigated the electron-phonon coupling (EPC) parameter $\lambda$, the logarithmic average phonon frequency $\omega_{log}$, and the Eliashberg phonon spectral function $\alpha^2F(\omega)$ to explore the possible superconductivity for the phase $P6_3/mmc$ PoH, *Pnma* $PoH_2$, $C2/c$ $PoH_4$ and $C2/m$ $PoH_6$ at different pressures. The obtained EPC parameter $\lambda$ is 0.33 for PoH (at 300 GPa), 0.19 for $PoH_2$ (at 200GPa), 1.09 for $PoH_4$ (at 200 GPa) and 0.43 for $PoH_6$ (at 200 GPa), while the calculated $\omega_{log}$ from the phonon spectrum reach 806.1, 712.9, 602.6 and 836.1 K, respectively. The superconducting critical temperature $T_c$ of Po-H system stable phases can be estimated by using the Allen-Dynes-modified McMillan equation[30]

$$T_C = \frac{\omega_{\log}}{1.2}\exp\left[\frac{1.04(1+\lambda)}{\lambda-\mu^*(1+0.62\lambda)}\right]$$ with the calculated logarithmic average frequency $\omega_{\log}$ and the Coulomb pseudopotential $\mu^*$ is often taken as ~ 0.1 and 0.13. The measured superconducting transition temperature $T_c$ values for *P*$6_3$/*mmc* PoH, *Pnma* $PoH_2$, *C*2/*c* $PoH_4$ and *C*2/*m* $PoH_6$ are 0.14-0.65 K (at 300 GPa), 0 K(at 200 GPa), 41.1-47.2K (at 200 GPa) and 2.25-4.68 K (at 200 GPa), respectively.

Figure 4 shows the evaluated Eliashberg phonon spectral function $\alpha^2F(\omega)$ and the partial electron-phonon integral $\lambda(\omega)$ for *P*$6_3$/*mmc* PoH, *C*2/*c* $PoH_4$ and *C*2/*m* $PoH_6$ to understand the origin of $\lambda$. Taken altogether, the electron–phonon integral has two regions from low to high frequency in the above structures. For *P*$6_3$/*mmc* structure of PoH at 300 GPa (Figure 4a), the low-frequency mode below 10 THz, contributing 46 % of the total value $\lambda$, comes from Po vibrations, whereas high frequency mode above 36 THz associated with H atoms provide a contribution 54 % in total EPC parameter $\lambda$. For the *C*2/*c* of $PoH_4$ at 200 GPa (Figure 4b), Po vibrations in the frequency region below 50 THz contribute approximately 39.3 % in total $\lambda$, and the remaining 60.7 % is from the H vibrations. *C*2/*m* $PoH_6$ was found that the Po and H vibration modes account for 25.5 % and 75.5 % of $\lambda$ at 200 GPa (Figure 4c).

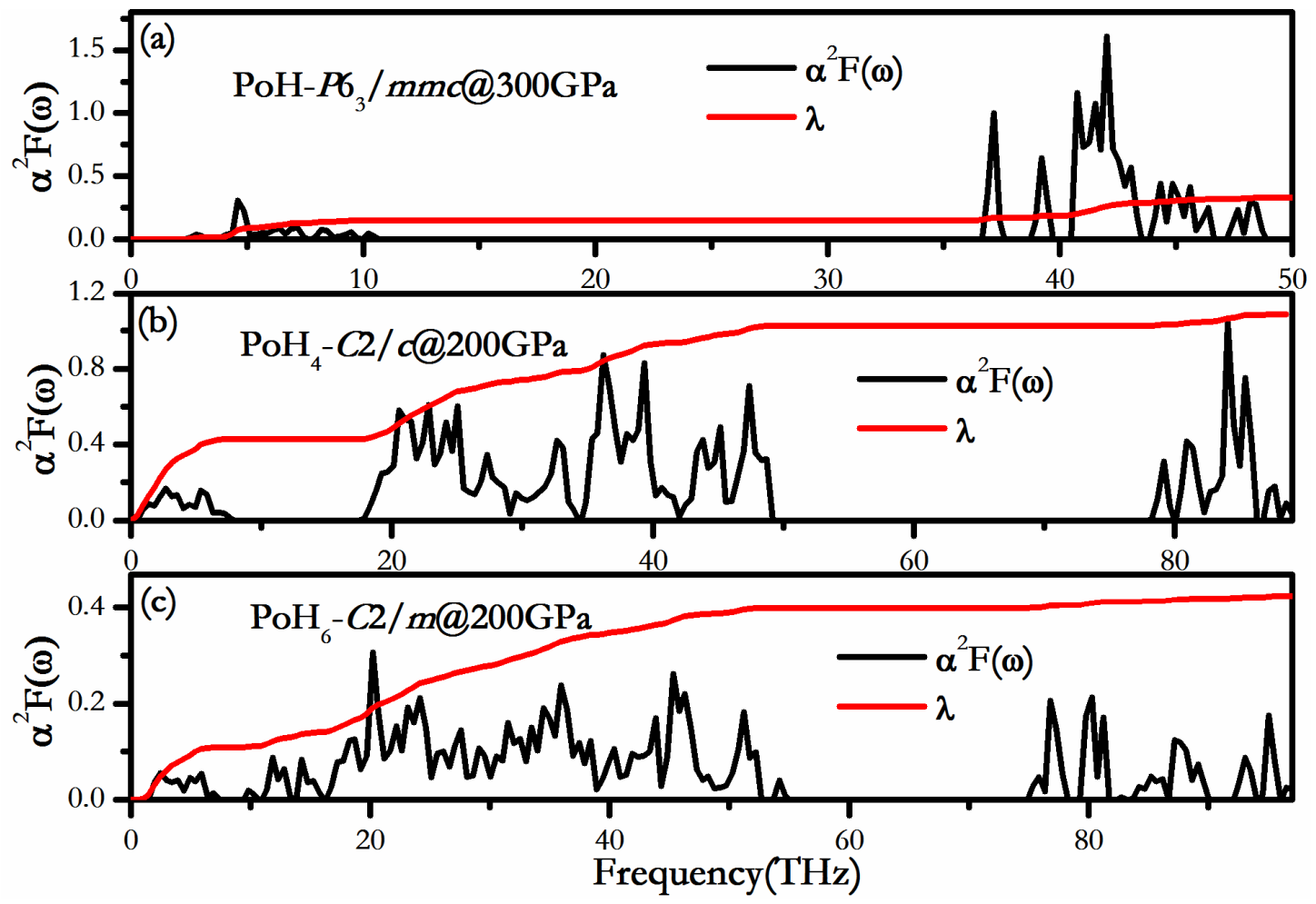

**Figure 4** The Eliashberg phonon spectral function $\alpha^2F(\omega)$ and the partial electron-phonon integration $\lambda(\omega)$: (a) PoH-*P*$6_3$/*mmc* at 300 GPa, (b) $PoH_4$-*C*2/*c* at 200 GPa and (c) $PoH_6$-*C*2/*m* at 200 GPa.

## Conclusion

In summary, we explore the structures, phase diagram and superconductivity of polonium hydrides under pressure using *ab initio* calculations. The polonium hydrides do not become thermodynamically stable with respect to decomposition into the Po and H up to 84.5 GPa. Upon further compression and in the excess hydrogen environment, PoH, $PoH_2$, $PoH_4$ and $PoH_6$ hydrides become thermodynamically preferred. Except PoH $P6_3/mmc$, a remarkable feature of the predicted stable structures is the presence of $H_2$ units. Electron-phonon coupling calculations show that the $T_c$ values of $P6_3/mmc$ PoH, $C2/c$ $PoH_4$, and $C2/m$ $PoH_6$ are 0.14-0.65 K (at 300 GPa), 41.1-47.2K (200 GPa) and 2.25-4.68 K (at 200 GPa), respectively. Our results provide a better understanding of the pressure-induced phase diagram and superconductivity of Po-H system, which has major implications for investigating other hydrides under pressure.

## Acknowledgments

This work was supported by the National Basic Research Program of China (No. 2011CB808200), Program for Changjiang Scholars and Innovative Research Team in University (No. IRT1132), National Natural Science Foundation of China (Nos. 11204100, 51032001, 11074090, 10979001, 51025206, 11104102, and 11404134), National Found for Fostering Talents of basic Science (No. J1103202), China Postdoctoral Science Foundation (2012M511326, 2013T60314, and 2014M561279). Parts of calculations were performed in the High Performance Computing Center (HPCC) of Jilin University.

# Supplementary information

# Phase diagram and properties of polonium hydrides under high pressure

Yunxian Liu, Defang Duan, Fubo Tian, Da Li, Xiaojing Sha, Zhonglong Zhao, Huadi Zhang, Gang Wu, Hongyu Yu, Bingbing Liu, & Tian Cui

*State Key Laboratory of Superhard Materials, College of Physics, Jilin University, Changchun 130012, People's Republic of China*

*Corresponding Author: E-mail: cuitian@jlu.edu.cn

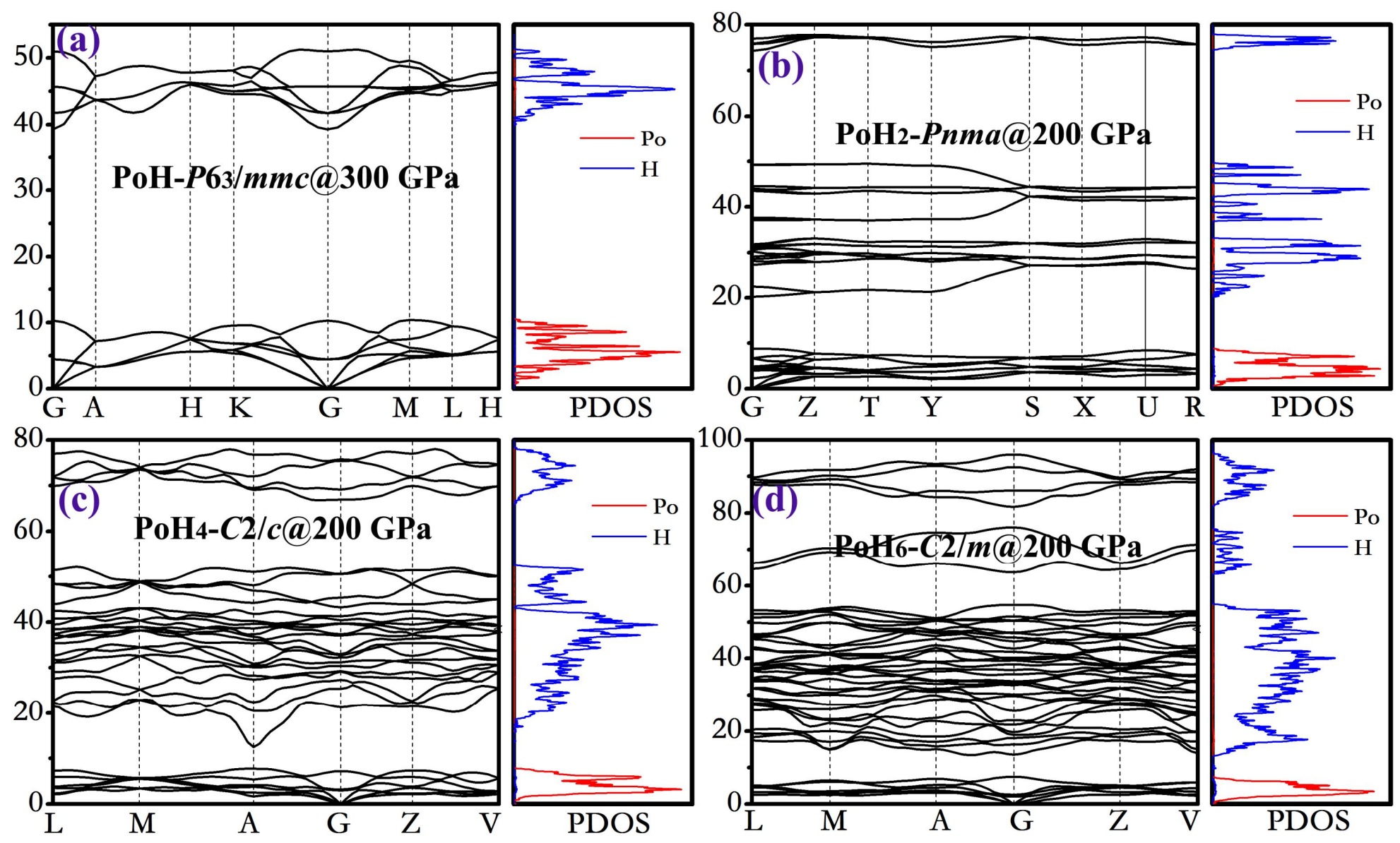

**Figure S1:** (a-d) The phonon band structure and projected phonon DOS charts for PoH-$P6_3/mmc$, $PoH_2$-$Pnma$, $PoH_4$-$C2/c$ and $PoH_6$-$C2/m$ at different pressure, respectively.

**Table S1:** Structural parameters of our predicted stable structures of Po-H system at selected pressure.

| Space group<br>Pressure | Lattice parameters<br>(Å, °) | Atomic coordinates (fractional) | | | | Sites |
|---|---|---|---|---|---|---|
| PoH-$P6_3/mmc$ | a= b=2.84590 | H1 | 0.00000 | 0.00000 | 0.50000 | 2a |
| 300 GPa | c=4.83120 | Po1 | 0.33333 | 0.66667 | 0.25000 | 2c |
| | α=β= 90 | | | | | |
| | γ=120 | | | | | |
| $PoH_2$-*Cmcm* | a=3.24840 | H1 | 0.12213 | 0.54851 | 0.75000 | 8g |
| 50 GPa | b=9.85620 | Po1 | 0.50000 | 0.65383 | 1.25000 | 4c |
| | c=3.39480 | | | | | |
| | α=β=γ=90 | | | | | |
| $PoH_2$-*Pnma* | a=5.26800 | H1 | 0.51807 | 0.25000 | 0.02754 | 4c |
| 300 GPa | b=2.82900 | H5 | 0.45813 | 0.75000 | 0.79165 | 4c |
| | c=4.44940 | Po1 | 0.18676 | 0.25000 | 0.86140 | 4c |
| | α=β=γ=90 | | | | | |
| $PoH_4$-$C2/c$ | a=4.07630 | H1 | -0.63532 | -0.18961 | -0.05386 | 8f |
| 200 GPa | b=4.08450 | H2 | -0.16518 | 0.05410 | -0.90263 | 8f |
| | c=7.89230 | Po1 | -0.50000 | 0.10458 | -0.75000 | 4e |
| | α=γ=90 | | | | | |
| | β=138.1964 | | | | | |
| $PoH_6$-$C2/m$ | a=12.13570 | H1 | -0.53080 | 0.35162 | 0.25074 | 8j |
| 200 GPa | b=2.79620 | H5 | -0.78661 | 0.00000 | 0.83013 | 4i |
| | c=2.96330 | H7 | -0.74323 | 0.00000 | 0.15901 | 4i |
| | α=γ=90 | H9 | -0.15273 | 0.00000 | 0.28030 | 4i |
| | β=90.2911 | H11 | -0.24763 | 0.00000 | 0.57331 | 4i |
| | | Po1 | -0.61084 | 0.00000 | 0.75290 | 4i |